\pdfoutput=1 
\documentclass[twocolumn,aps,prl,superscriptaddress]{revtex4-2}
\usepackage[T1]{fontenc}

\usepackage{graphicx}
\usepackage{subfigure}
\usepackage{epsfig}
\usepackage{amssymb,amsmath,amsopn,amsfonts}
\usepackage{colordvi}
\usepackage{color}
\usepackage{subeqnarray} 
\usepackage{bbold}
\usepackage{xcolor} 
\usepackage[symbol]{footmisc} 

\newcommand{\avg}[1]{\langle {#1} \rangle}

\newcommand{\pt}{\mathcal{PT}}

\newcommand{\transpose}{\mathrm{T}}
\newcommand{\ket}[1]{\left | #1 \right \rangle}
\newcommand{\bra}[1]{\left \langle #1 \right |}
\newcommand{\braket}[2]{\left \langle #1 | #2 \right \rangle}

\newcommand{\proj}[1]{\ket{#1}\!\!\bra{#1}}
\newcommand{\ttt}[1]{\text{#1}}

\newcommand{\gammaC}{\gamma_{\ttt{c}}}

\newcommand{\Appendix}{Appendix}
\newcommand{\ketNull}{\ket{\ttt{vac}}}

\newcommand{\NobuB}{Currently at: Department of Communications Engineering, Graduate School of Engineering, Tohoku University, Sendai, Japan}
\newcommand{\NobuA}{NTT Basic Research Laboratories, NTT Corporation, Atsugi, Japan}
\newcommand{\Levon}{Present address: Krisp, Hrachya Kochar 5/1, Yerevan 0033, Armenia}

\newcommand{\Jack}{Center for Hybrid Quantum Networks (Hy-Q), Niels Bohr Institute, University of Copenhagen, Blegdamsvej 17, DK-2100 Copenhagen, Denmark}
\newcommand{\hashi}{  NTT Device Technology Laboratories, NTT Corporation, Atsugi, Japan}

\begin{document}
 
\title{Photonic quantum simulations of coupled $\mathcal{PT}$-symmetric Hamiltonians}

\author{Nicola~Maraviglia}
\affiliation{Quantum Engineering Technology Labs, H. H. Wills Physics Laboratory and Department of Electrical and Electronic Engineering, University of Bristol, Bristol BS8 1FD, UK}
\author{Patrick~Yard}
\affiliation{Quantum Engineering Technology Labs, H. H. Wills Physics Laboratory and Department of Electrical and Electronic Engineering, University of Bristol, Bristol BS8 1FD, UK}
\author{Ross~Wakefield}
\affiliation{Quantum Engineering Technology Labs, H. H. Wills Physics Laboratory and Department of Electrical and Electronic Engineering, University of Bristol, Bristol BS8 1FD, UK}
\author{Jacques~Carolan}
\affiliation{\Jack}
\author{Chris~Sparrow}
\affiliation{Quantum Engineering Technology Labs, H. H. Wills Physics Laboratory and Department of Electrical and Electronic Engineering, University of Bristol, Bristol BS8 1FD, UK}
\affiliation{Department of Physics, Imperial College London, SW7 2AZ, UK}
\author{Levon~Chakhmakhchyan}
\affiliation{Quantum Engineering Technology Labs, H. H. Wills Physics Laboratory and Department of Electrical and Electronic Engineering, University of Bristol, Bristol BS8 1FD, UK}
\affiliation{\Levon}
\author{Chris Harrold}
\affiliation{Quantum Engineering Technology Labs, H. H. Wills Physics Laboratory and Department of Electrical and Electronic Engineering, University of Bristol, Bristol BS8 1FD, UK}
\author{Toshikazu~Hashimoto}
\affiliation{\hashi}
\author{Nobuyuki~Matsuda}
\affiliation{\NobuA}
\affiliation{\NobuB}
\author{Andrew~K.~Harter}
\affiliation{Department of Physics, Indiana University Purdue University Indianapolis (IUPUI), Indianapolis, Indiana 46202, USA}
\author{Yogesh~N.~Joglekar}
\email[]{yojoglek@iupui.edu}
\affiliation{Department of Physics, Indiana University Purdue University Indianapolis (IUPUI), Indianapolis, Indiana 46202, USA}
\author{Anthony~Laing}
\email[]{anthony.laing@bristol.ac.uk}
\affiliation{Quantum Engineering Technology Labs, H. H. Wills Physics Laboratory and Department of Electrical and Electronic Engineering, University of Bristol, Bristol BS8 1FD, UK}

\begin{abstract}
Parity-time ($\pt$) symmetric Hamiltonians are generally non-Hermitian
and give rise to exotic behaviour in quantum systems at exceptional points,
where eigenvectors coalesce.
The recent realisation of $\pt$-symmetric Hamiltonians in quantum systems
has ignited efforts to simulate and investigate many-particle quantum systems across exceptional points.
Here we use a programmable integrated photonic chip to simulate a model
comprised of twin pairs of $\pt$-symmetric Hamiltonians,
with each the time reverse of its twin.
We simulate quantum dynamics across exceptional points
including two- and three-particle interference,
and a particle-trembling behaviour that arises due to interference between subsystems undergoing time-reversed evolutions.
These results show how programmable quantum simulators can be used to investigate foundational questions in quantum mechanics.
\end{abstract}
\maketitle

Dirac Hermiticity of a Hamiltonian has been a tenet of quantum theory since its inception. This constraint guarantees real eigenvalues, orthogonal eigenstates, and a unitary time evolution; it reflects the dynamics of an isolated system.  Over the past two decades, non-Hermitian Hamiltonians that are symmetric under combined parity ($\mathcal{P}$) and time-reversal ($\mathcal{T}$) transformations were extensively investigated, first mathematically~\cite{Bender2007,AM2010} and then experimentally in classical wave systems~\cite{ElGanainy2018,Miri2019}. In the classical domain, $\mathcal{PT}$-symmetric systems have shown remarkable properties such as unidirectional invisibility~\cite{Feng2012}, single-mode~\cite{Feng2014,Hodaei2014} and topological~\cite{ Harari2018,Bandres2018} lasing, topological energy transfer~\cite{Xu2016},  mode switching~\cite{Doppler2016}, and enhanced sensitivity~\cite{Lai2019,Hokmabadi2019}. While generally non-Hermitian, $\mathcal{PT}$-symmetric Hamiltonians exhibit exceptional points (EPs) at which two (or more) eigenvalues become degenerate and the corresponding eigenstates coalesce. For all these advances, the role of thermal and quantum noise on EPs in these semi-classical systems is not yet understood~\cite{Wang2020,Wiersig2020}. 

Realizing $\mathcal{PT}$-symmetric systems in the quantum domain has been a major challenge~\cite{Scheel2018-NoGain}, and its circumvention has relied on two methods. The first approach, known as passive $\pt$ symmetry, is based on the equivalence (after normalisation) of a $\pt$-symmetric Hamiltonian and a dissipative Hamiltonian with mode-selective losses~\cite{Xiao2017-topological,Li2016-floquet}. Such a dissipative Hamiltonian arises naturally from a Lindblad approach for open quantum systems when certain quantum jumps are ignored and allows the implementation of coherent, non-unitary dynamics in a minimal quantum system~\cite{Naghiloo2019-expSupCond}. The second method uses Hamiltonian dilation that places the $\pt$-symmetric Hamiltonian in a larger Hilbert space by introducing an ancillary two-level system~\cite{Gunther2008-proposal,Xiao2018-experimentOptic,Wu2018-PTinSpin}, and the coherent, non-unitary dynamics of interest are recovered from the unitary dynamics in the larger Hilbert space. In the first approach, the evolution time of the system is limited by the exponentially decaying signal or post-selection success probability, while the second method requires a time-dependent Hermitian Hamiltonian for the larger Hilbert space.

Here we propose a framework for the quantum simulation of $\pt$-symmetric Hamiltonians that is particularly suited to technology platforms that allow the direct implementation of unitary transformations \cite{Lau2012-UniversalIons,Shen2014-UnitaryIons,Xiu2017-Microwaves}. The non-unitary evolution operator, and a second operator that is the time reverse of the first, are embedded together into a global unitary transformation that is then implemented by the device. The overall evolution in this model allows single- or many-particle excitations to tunnel between the twin systems, with a probability that scales with the non-Hermicity of the simulated Hamiltonians. This construction permits the experimental investigation of states superposed across opposite directions of time in the context of non-Hermitian Hamiltonians.

We experimentally simulate multiparticle dynamics~\cite{Klauck2019} in two- and three-mode $\pt$-symmetric Hamiltonians
using a programmable photonic chip \cite{carolan2015-universal},and ensembles of one, two, and three photon input states.
We reproduce the dynamics in the $\pt$-symmetric regime (defined by real eigenvalues) and across the EP into the $\mathcal{PT}$-symmetry broken regime (defined by complex conjugate eigenvalues), including the effects of mutual coherence and interference between the forward-in-time and backward-in-time subsystems. Combined with the current advances in photonic technologies, these demonstrations point towards the inclusion of non-Hermitian simulations into the realm of quantum technologies.


\subsection{Model and simulation procedure}




Our simulation approach begins by considering a family of $\pt$-symmetric systems consisting of a linear chain of $N$ modes with a coupling amplitude $J>0$ between nearest-neighbour modes. In addition, an imaginary potential of opposite sign $\pm i\gamma$ located in the two spatially symmetric end modes determines a gain and a loss site.  We fix the energy scale of our system by choosing $J=+1$ so that the Hamiltonian reads
\begin{align}
\label{eq:Nchain}
H_N(\gamma)
= - \sum_{k=1}^{N-1} (\ket{k}\!\!\bra{k+1} + \ket{k+1}\!\!\bra{k}& ) \nonumber \\
+ i\gamma (\proj{1} & - \proj{N})
\end{align}
where $\ket{k}$ is the state associated to the $k$-th mode. Using $\{  \ket{k}\}_{k=1}^N$ as a basis, the parity operator we choose is represented by the antidiagonal matrix $\mathcal{P}_{mn} = \delta_{m, N+1-n}$ and the time-reversal operator $\mathcal{T}$ corresponds to the complex conjugation operation. $H_N(\gamma)$ presents an EP for a critical value $\gamma=\gammaC$ that depends on $N$ \cite{Jin2009-Nmodes,Joglekar2010}. The spectrum of the Hamiltonian $H_N(\gamma)$ is purely real for $\gamma<\gammaC$ and the corresponding eigenvectors are simultaneous eigenvectors of the antilinear $\mathcal{PT}$ operator (unbroken symmetry phase). At the EP, $\gamma=\gammaC$, two or more states coalesce, while at $\gamma>\gammaC$, two eigenstates with opposite imaginary eigenvalue emerge from the previously coalescing states (broken symmetry phase). The full spectrum of the Hamiltonian can be obtained by solving a set of transcendental equations reported in \cite{Jin2009-Nmodes} but $\gammaC$ is given by a simple closed form:  $\gammaC/J=1$ whenever $N$ is even and  $\gammaC/J =\sqrt{(N+1)/(N-1)}$ when $N$ is odd. For a static, non-Hermitian Hamiltonian $H_N(\gamma)$, the non-unitary time evolution operator $G_{N}(\gamma, t)$ is given by (assuming $\hbar=1$) 
\begin{equation}
G_{N}(\gamma, t)=\exp\left[-i H_{N}(\gamma)\,t\right]. 
\end{equation}
We note that these choices set the time-scale for our system, $\hbar/J$, to unity. 

\begin{figure*}
	\centering
	\includegraphics[trim=0 0 0 0,width=\textwidth]{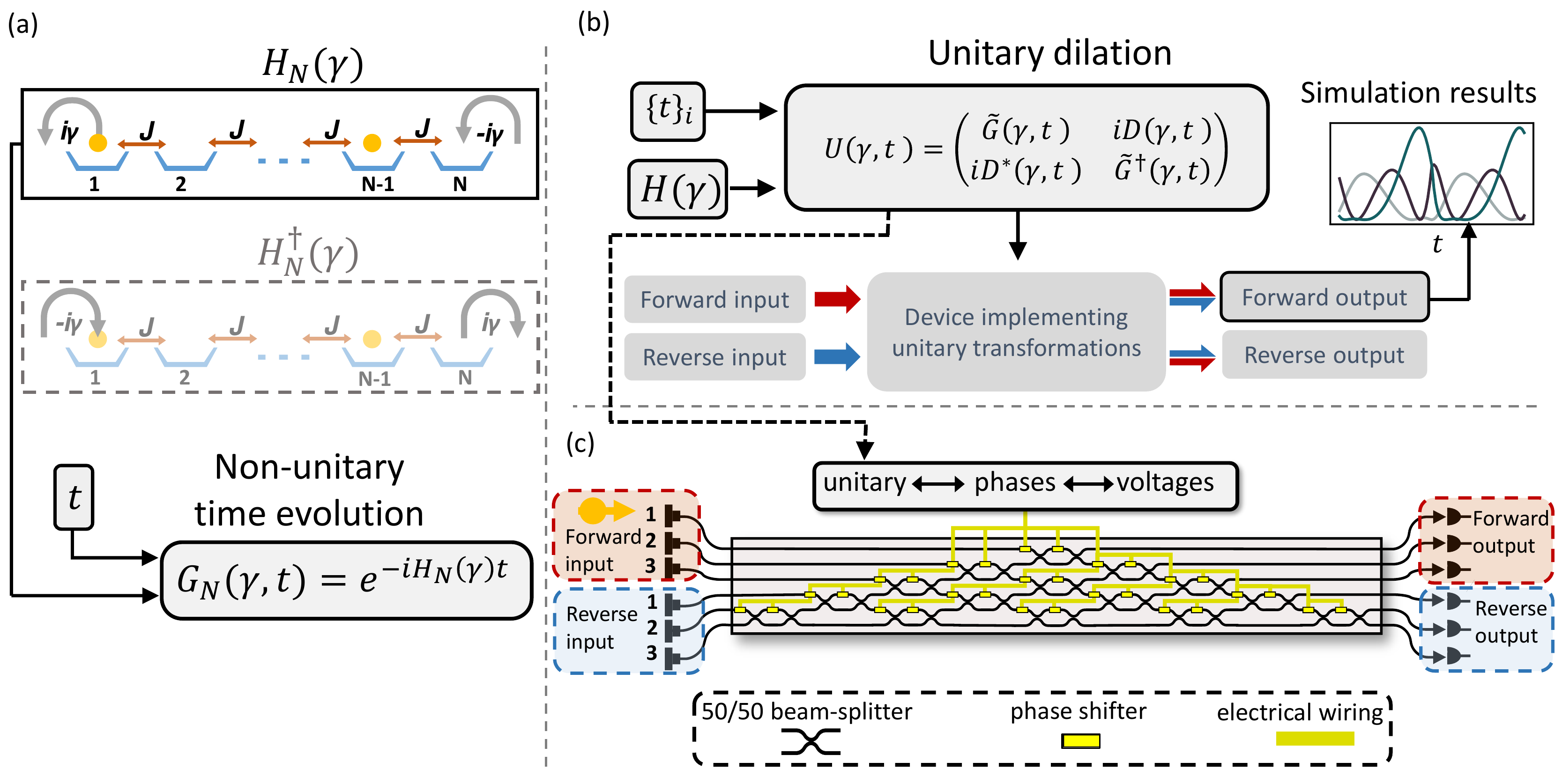}
	\caption{
		(a) Graphical representation of the simulated $\pt$-symmetric non-Hermitian Hamiltonian $H_N(\gamma)$ that describes an open quantum system. Excitations, represented by yellow circles, can populate the $N$ modes of a linear chain. 
  		Nearest neighbour modes are coupled by a potential  $J$ that is set to unity during the simulations reported in this work.  
		The imaginary potential, $\pm i \gamma$, has the effect of amplifying and deamplifying the probability amplitude of an excitation in the first and last mode, respectively. 
		$ H_N^\dag(\gamma)  $ is equivalent to $ H_N(\gamma)$ with the first and last modes swapped. 
		The time evolution of a system with Hamiltonian $H_N(\gamma)$ is described by a non-unitary operator $G_{N}(\gamma,t)$
		(b) Simulation procedure based on unitary dilation. Given a non-Hermitian Hamiltonian $H(\gamma)$ and an evolution time $t$, it is possible to create a 
		unitary $U(\gamma,t)$ that consists of  $\tilde{G}(\gamma,t)=G(\gamma,t)/||G(\gamma,t)||$ and an auxiliary matrix $D(\gamma,t)$.  
		The simulation proceeds by implementing $U(\gamma,t)$ and initialising a desired input state in modes of the device corresponding to the forward system input. 
		The effect of the unitary device will transform the input and distribute it across the forward and reverse output modes. 
		When the output state is detected in the forward output, the input state successfully transformed via $\tilde{G}(\gamma,t)$. 
		(c) Photonic implementation of the quantum simulation. Single photon states are injected into the input ports of the first $N=3$ waveguides of a configurable integrated photonic circuit. Forward and reverse output are detected at the first and last output ports of the interferometer.     
		A triangular mesh of integrated beam-splitters and metallic phaseshifters allows us to reconfigure the integrated interferometer to implement the desired $U(\gamma,t)$.	 
	}
	\label{fig:fig1}
\end{figure*}

To simulate its dynamical evolution with a device that implements unitary transformations, we adopt the Halmos unitary dilation~\cite{halmos-1950,Wang:20-DirectGraph}. Namely, we first rescale $G_N(\gamma,t)$ by its largest singular value $||G_{N}(\gamma,t)||_2 $ obtaining
\begin{equation}
\tilde{G}_{N}(\gamma,t):= G_{N}(\gamma,t)/||G_{N}(\gamma,t)||_2,
\end{equation}
so that the operator norm  of the scaled operator $\tilde{G}_N(\gamma,t)$ is unity. Then, defining the defect operator as 
\begin{equation}
D_N(\tilde{G}_N):=\left[{\mathbb 1}-\tilde{G}_{N}(\gamma,t) \tilde{G}_{N}^\dagger(\gamma,t)\right]^{1/2},
\end{equation}
we construct the following unitary transformation $U_{2 N}(\gamma,t)$ defined on a Hilbert space twice the size. The block matrix representation of $U_{2 N}(\gamma,t)$ is  

\begin{equation}
U_{2N}(\gamma,t)=\begin{bmatrix}
\tilde{G}_N(\gamma,t) & iD_N(\tilde{G}_N) \\
iD_N(\tilde{G}_N^\dagger) & \tilde{G}^\dagger_N(\gamma,t)
\end{bmatrix}.
\end{equation}

We map the modes of the dilated Hilbert space onto the spatial modes of a reprogrammable linear optical chip. The time evolution of excitations in the $\pt$-symmetric system is simulated by injecting single photons into the optical circuit and reprogramming it to implement the unitary transfer matrix $U_{2 N}(\gamma,t)$ for a sequence of time steps $\{U_{2 N}(\gamma,t_i)\}_i$, see Fig.~\ref{fig:fig1}. To simulate the evolution determined by $H_N(\gamma)$, we encode the initial state in the first $N$ modes of the interferometer and consider detection events on the outputs of those same modes. Our construction is applicable for $\mathcal{PT}$-symmetric systems even when the spectrum of the Hamiltonian becomes complex, allowing simulations across exceptional points. We also emphasize that since we realize the unitary, and not the (dilated) Hamiltonian, time is a parameter and so we can simulate arbitrary time scales. Specifically, we can simulate the long time dynamics expected at the EP, due to the vanishing eigenvalue gap. 

Interestingly, since the Hamiltonian is transpose-symmetric, i.e. $H_N^\dagger(\gamma)=H_N^*(\gamma)$, the global unitary $U_{2N}(\gamma,t)$ describes two coupled $\pt$-symmetric systems that are the time reverse of each other. Indeed, $\tilde{G}^\dagger_N(\gamma,t)=\tilde{G}^*_N(\gamma,t)=\mathcal{T}\tilde{G}_N(\gamma,t)\mathcal{T}^{-1}$, meaning that a system initialised in the last $N$ modes of the interferometer transforms with a backwards-in-time (reverse) evolution. 
Whenever $\gamma \neq 0$, the two systems are coupled by the defect matrix $D_N$ and excitations can tunnel between these two systems. In addition to simulating individual  $\pt$ systems, the statistics output from the photonic chip track the unitary evolution of the closed composite system.

We also note that the above dilation process is not specific to $\pt$-Hamiltonians. Any non-Unitary evolution operator, $ G_{N}(t)$, can be embedded in a larger unitary transformation, following the above method. This process has already been used to simulate Lindbladian dynamics \cite{Sparrow2018-Nature} and could be extended to simulate anti-$\pt$ systems \cite{peng2016anti}. 
\begin{figure*}
	\centering
	\includegraphics[trim=0 0 0 0,clip=true,width=\textwidth]{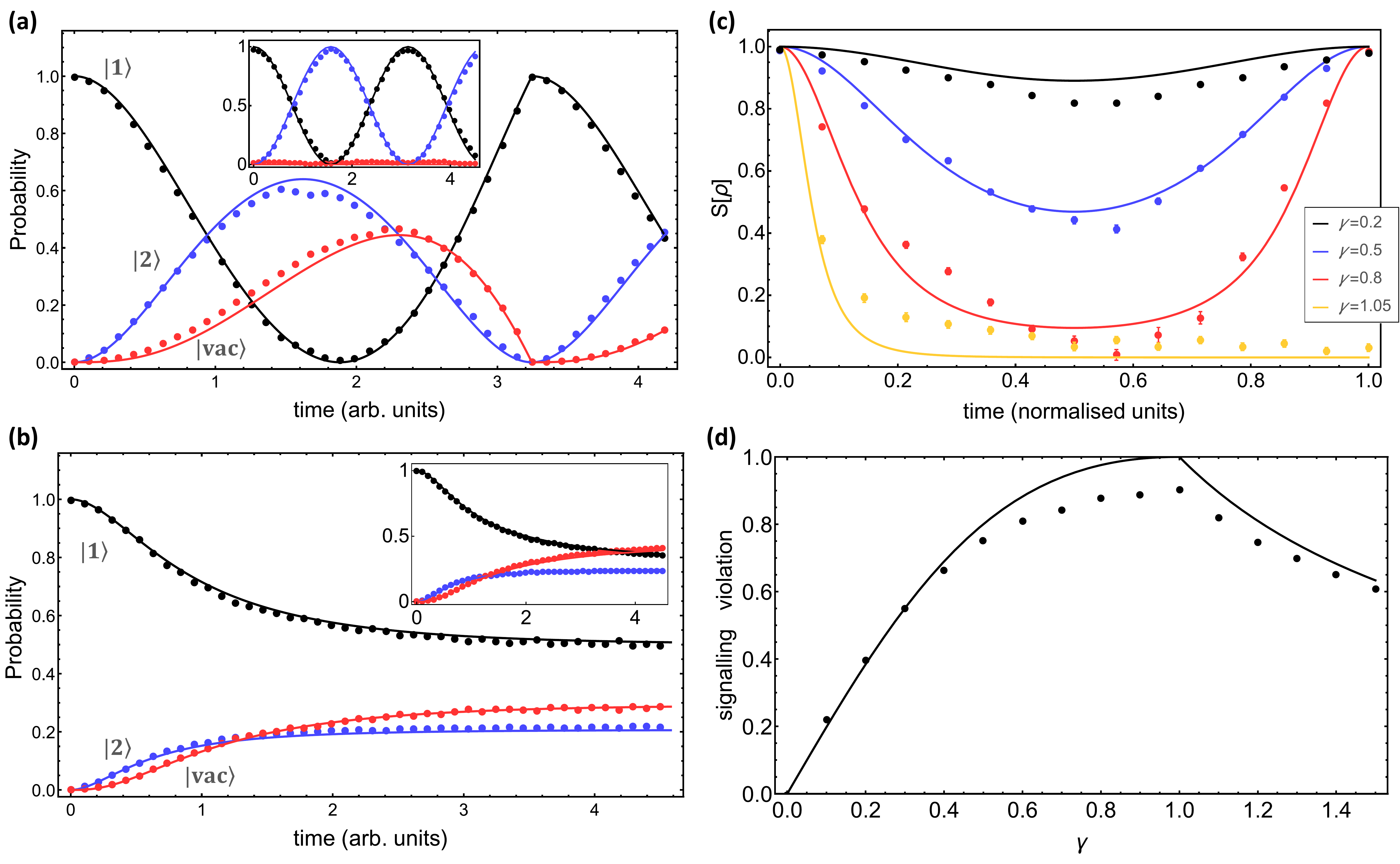}
	\caption{
	Simulation of the forward system across different regimes of the $\pt$-Hamiltonian. 
	Solid lines (dots) correspond to the theory (experimental data).
	(a,b)
	A single excitation is initialised in $\ket{1}_\text{F}$ and observed in the $\left\{\ketNull,\ket{1},\ket{2}\right\}_\text{F}$ basis as shown with red, black and blue data, respectively.
	(a) In the unbroken symmetry phase ($\gamma=0.25$),
	the plot shows the tunnelling of excitations between the forward and reverse systems
	resulting in a periodic oscillation in the vacuum probability;
	the inset shows the Hermitian regime ($\gamma=0$)
	with a constant vacuum probability of $0$,
	and sinusoidal oscillations between the two populated levels.
	(b) In the broken symmetry phase ($\gamma=1.1$),
	the evolution is not periodic and the system tends rapidly to a steady state;
	the inset shows the evolution of the system at the exceptional point.
	(c,d) Evolution of the initial state $\rho_0=0.5\ket{1}_\ttt{F}\!\bra{1}_\ttt{F}+0.5\ket{2}_\ttt{F}\!\bra{2}_\ttt{F}$. 
	(c)	The entropy of the system is periodic for $\gamma<1$ and goes asymptotically to 0 for $\gamma>1$. Time is reported in units normalised as a function of $\gamma$ by dividing by $\tau=\pi/\sqrt{|1-\gamma^2|}$.
	(d) Signalling violation as a function of $\gamma$ and for evolution time $t=\pi/\sqrt{|1-\gamma^2|}$ with a maximum at the exceptional point $\gamma=1$. The outcomes of measurements on one qubit are affected by local operations on a $\pt$ qubit, entangled with the first.
	}
	\label{fig:fig2}
\end{figure*}


\subsection{Experimental procedure}
Our photonic chip comprises a triangular mesh of 15 interconnected Mach-Zehnder interferometers,
controlled by thirty metallic thermo-optic phase shifters,
and is capable of implementing any unitary linear optical transformation over its six waveguides 
(see Ref. \cite{carolan2015-universal} for additional device details). 
By means of the algorithm demonstrated by Reck et al. for the decomposition of any discrete optical unitary operator \cite{Reck-universal},  for each $U_{2 N}(\gamma,t_i)$ we retrieve the list of phases which, when applied in our interferometer, implement the desired unitary transformation. Calibration of the on-chip thermo-optic phase shifters allows us to map a desired phase shift to a voltage bias to be set.
Input photons are generated with a bulk spontaneous parametric down-conversion source,
and coupled in and out of the chip with packaged optical fibres,
before detection with silicon avalanche photo-diodes.
This setup is capable of simulating coupled $\pt$-symmetric Hamiltonians with $N=2$ and $N=3$ modes in each subsystem and multi-particle evolution. 

Tuning the strength of the complex potential $\gamma$ corresponds to changing the unitary matrix $U_{2 N}(\gamma,t_i)$, and therefore the set of phases dialled on our chip. In this way, all values of $\gamma$ can be implemented in the same way, allowing us to investigate the system dynamics in four regimes: uncoupled Hermitian evolution, coupled $\pt$-symmetric evolution, evolution at the EP, and evolution in the broken symmetry phase. As we increase $\gamma$ we also increase the coupling between the systems, $iD(\gamma,t)$, which correspondingly decreases the probability of a photon remaining in the desired subsystem.


To distinguish between the state of the simulated system
and the photonic Fock state of our simulator,
we will use the subscript $p$ for the latter.
That is, for $N=2$, Fock states $\{\ket{1_p0_p}, \ket{0_p1_p}\}$,
respectively, encode the states $\{\ket{1}, \ket{2}\}$,
while for $N=3$, Fock states $\{\ket{1_p0_p0_p}, \ket{0_p1_p0_p}, \ket{0_p0_p1_p}\}$
 encode the states $\{\ket{1}, \ket{2}, \ket{3}\}$.
The Fock states
$\ket{0_p0_p0_p}$ or $\ket{0_p0_p}$,
corresponding to an unoccupied system,
will be represented with $\ketNull$.
Furthermore, subscript F or R will differentiate
between the forward and reverse systems,
or equivalently, between the 
first or last $N$ modes of our interferometer, respectively.


\subsection{Simulation of two-mode $\pt$ symmetric systems}

\begin{figure*}[t!]
	\centering
	\includegraphics[trim=0 0 0 0,width=1.0\textwidth]{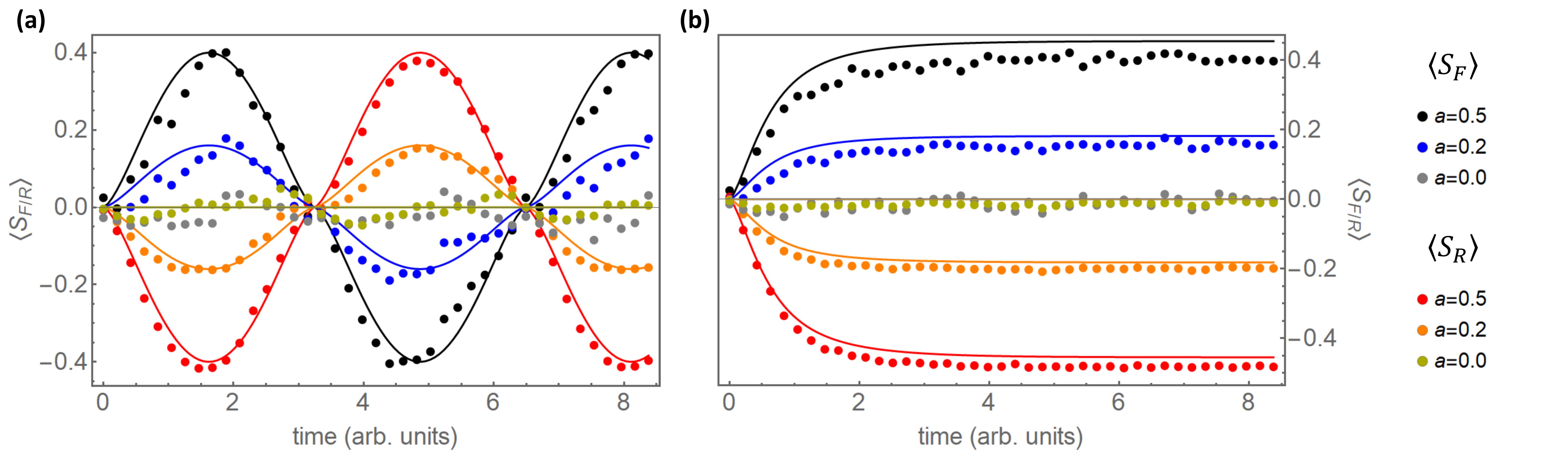}	
	\caption{
	Simulation of Zitterbewegung-like interference effects between coupled, time-reversed, two-mode $\pt$-systems.
	Expectation value of $S_\ttt{F(R)} =\ket{1}_\ttt{F(R)}\bra{2}_\ttt{F(R)}+\ket{2}_\ttt{F(R)}\bra{1}_\ttt{F(R)} $
	for the system initialised as 
	${\rho_\ttt{in}=1/2(\ket{1}_\ttt{F}\bra{1}_\ttt{F}+\ket{1}_\ttt{R}\bra{1}_\ttt{R})+a(\ket{1}_\ttt{F}\bra{1}_\ttt{R}+\ket{1}_\ttt{R}\bra{1}_\ttt{F})}$ are shown. 
	The parameter $a$ characterises the degree of coherence between the time-reversed systems.
	(a) In the unbroken symmetry phase ($\gamma=0.25$),
	the amplitude of the oscillations of $\avg{S_\ttt{F(R)}}$ are proportional to the initial level of coherence of the input state.
	(b) In the broken symmetry phase ($\gamma=1.1$),
	the initial level of coherence of the input state dictates the saturation level of $\avg{S_\ttt{F(R)}}$.}
	\label{fig:fig4}
\end{figure*}

We first simulate 2-mode $\pt$-symmetric systems. With $N=2$, the eigenvalues of the Hamiltonian are $\epsilon=\pm\sqrt{1-\gamma^2}$. Since the global model in this instance comprises 4 modes in total, there is sufficient redundancy in the six-mode chip to implement a separate tomography stage.
The first section of the chip implements the state evolution described by $\{U_{4}(\gamma,t_i)\}_i$.
The remaining part of the chip is used to perform 
projective measurements allowing the reconstruction of the density matrices for evolved single photon states. 

Figure~\ref{fig:fig2} shows the results from the simulations of the unbroken and broken symmetry phases. 
Experimental points are presented next to theory lines.
In Figs~\ref{fig:fig2}(a,b) we report the probability for an excitation initialised in the state $\ket{1 }_\ttt{F}\otimes\ketNull_\ttt{R}$ to be detected in any of the forward system modes but we enforce that the cumulative outcome statistic from the forward and reverse subsystems is normalised to one. 
To do so, we retained both the events where the propagating excitation was detected in the forward subsystem modes heralding its evolution according to $\tilde{G}$, and those where the excitation has left the forward modes to reappear in the reverse subsystem modes. 
If the excitation moves to the reverse system, the forward system is left in its vacuum state with no excitation, $\ketNull_\ttt{F}$, and as such is reported 
in Figs~\ref{fig:fig2}(a,b). A formal derivation of the forward state description as result of a partial trace over the modes of the reverse
 system is available in the \Appendix C. 
In the unbroken symmetry phase, shown in Fig.~\ref{fig:fig2}(a) with $\gamma=0.25$, 
the time-evolution of the system is periodic and
the finite probability of detecting $\ketNull_\ttt{F}$
indicates that the excitation can tunnel from the forward system to its time reversed twin.
In contrast, the inset in the same plot shows the evolution when $\gamma=0$.
Since the system is closed and Hermitian, only sinusoidal oscillations of the probabilities of detecting $\ket{1 }_\ttt{F}$ and $\ket{2 }_\ttt{F}$ are observed.
Figure~\ref{fig:fig2}(b) shows, instead, the evolution in the broken symmetry phase with $\gamma=1.1$.
The system rapidly tends to a steady state
with no oscillations and with a non-zero probability of finding the excitation in either the forward or reverse systems.
Such a qualitative distinction between the two regimes is caused by the characteristic phase transition
at the EP.
In the inset of Fig.~\ref{fig:fig2}(b), 
we report the evolution observed at the EP when $\gamma=1$. Here the period of the system oscillation  $\tau=\pi/\sqrt{1-\gamma^2}$  tends to infinity. The probabilities evolve towards a never-reached turning point and the systems effectively tends to a steady state with coefficients that follows a polynomial power law. As opposite, in the broken symmetry phase, the system approaches its steady state with an exponential power law whose characteristic time, $\tau'=\pi/\sqrt{\gamma^2-1}$,  decreases for increasing $\gamma$.  

Figures~\ref{fig:fig2}(c,d) show the evolution of a completely mixed input state, $\rho_0=0.5\ket{1}_\ttt{F}\!\bra{1}_\ttt{F}+0.5\ket{2}_\ttt{F}\!\bra{2}_\ttt{F}$, which was realised by first entangling two photons using a photonic gate \cite{Ou1988-bellState}.
The total unitary implemented by the chip is obtained by multiplying together the transfer matrix that describes the entangling gate, and $ U_{4}(\gamma,t)$, in such a way that one photon is injected into the 2 modes of the forward system of $U_{4}(\gamma,t)$ while the second photon of the entangled pair, is transmitted by two ancillary waveguides not affected by the $ U_{4}(\gamma,t) $. 
%
To herald the successful generation of the entangled photon pair, the tomographic reconstruction of the state is performed by retaining only the coincidental events of a photon detected in the forward $\pt$-symmetric system and a photon detected in the two ancillary waveguides. 

In Fig.~\ref{fig:fig2}(c) we report the evolution of the Von Neumann entropy $S[\rho(t)]$,  of the initially mixed $\pt$ qubit $\rho_0$, for different values of $\gamma$. As opposite to a closed Hermitian system whose entropy is a constant of motion, in the unbroken symmetry phase, the entropy is a non-monotonic function of time. By reporting the time parameter in units of  $\tau=\pi/\sqrt{|1-\gamma^2|} $, the entropy has a minimum when $t=0.5 \tau$. Instead, when $\gamma> \gammaC$,  the entropy decays to 0. 
The periodic behavior of the entropy in the $\mathcal{PT}$-symmetric phase indicates the flow of information from the forward system to its time-reversed twin and back~\cite{Ueda2017}. 

Figure ~\ref{fig:fig2}(d) shows the results of a signalling violation test, as described in \cite{Lee2014-NoSignaling,Tang2016-NoSignalling}. The test refers to the possiblity of affecting the state of a system $B$ by performing only local operation on a system $A$. For a pair of qubits, the signalling violation, $s.v.$, expressed in terms of the measurement probability $\ttt{P}$  on the qubit $B$, reads:
\begin{equation}
	s.v.=\Big[\ttt{P}(1|S)-\ttt{P}(2|S)\Big]-\Big[\ttt{P}(1|I)-\ttt{P}(2|I)\Big],
\end{equation}  
where $1( 2)$ corresponds to the first (second) level of the qubit $B$, and $I(S)$ refers to a local identity (swap) transformation applied to the qubit $A$.
Our simulation proceeds by entangling the qubits $A$ and $B$,
then we apply either the identity or the swap transformation,
and finally we simulate the evolution of qubit $A$ with the $\pt$-symmetric Hamiltonian.
Meanwhile, the qubit $B$ is encoded by the photon transmitted over the two ancillary waveguides. 
Figure~\ref{fig:fig2}(d) shows that the probabilities of the two alternative outputs for qubit $B$ are conditioned by the a local unitary transformation applied to the $\pt$ qubit $A$ prior to its non-unitary evolution. 
The no-signaling condition is violated for all $\gamma \neq 0$, and is maximal for $\gamma = \gammaC$. To maximise the signalling violation, the evolution time for the $\pt$ system is set to $t=\pi/\sqrt{|1-\gamma^2|}$.  

Since in our framework excitations initialised in one system naturally tunnel to its time-reversed twin,
we investigate the interference effects that arise from an initial excitation
coherently superposed across the forward and the reverse systems.
Interference between systems evolving in opposite directions in time
is an effect pertinent to the foundations of quantum mechanics~\cite{s-select},
and gives rise to peculiar quantum effects predicted in the dynamics of a relativistic electron
\cite{zitter1, klein1}, recently investigated with trapped ions and Bose-Einstein condensates~\cite{time_reverse3,time_reverse4}.

We choose to measure an observable that manifests a qualitative change depending both on the coherence between the two systems
and their deviation from hermiticity.
We define a generalisation of the $\sigma_x$ operator for the forward (reverse) system
\begin{equation}
S_\ttt{F(R)} = \begin{pmatrix}
0 & 0 & 0 \\
0 & 0 & 1 \\
0 & 1 & 0
\end{pmatrix},
\end{equation}
as represented in the
$\{\ketNull_\ttt{F(R)},\ket{1}_\ttt{F(R)},\ket{2}_\ttt{F(R)} \}$
basis.
Such an observable is a constant of motion in the Hermitian regime and can be considered as a discrete-variable equivalent of a particle coordinate operator~\cite{foot1}.
However, if we prepare an input state that is a coherent superposition of the forward and reverse subsystems, such as
$(\ket{1}_\ttt{F}+\ket{1}_\ttt{R})/\sqrt{2}$,
both $\avg{S_\ttt{F}}$ and $\avg{S_\ttt{R}}$
exhibit oscillatory behaviour.

The amplitude of these oscillations is reduced if the coherence between the states $\ket{1}_\ttt{F}$ and $\ket{1}_\ttt{R}$ decreases and it vanishes if the input is a statistical mixture of $\ket{1}_\ttt{F}$ and $\ket{1}_\ttt{R}$, Fig.~\ref{fig:fig4}(a).
We therefore associate this oscillatory behaviour with the coherent interference between forward and reverse modes of the $\pt$-symmetric doublet, a situation that is analogous to the Zitterbewegung effect \cite{zitter1}. This predicts an oscillatory or ``trembling" behaviour of the position expectation value for a relativistic electron, stemming from interference between positive and negative energy solutions of the Dirac equation.
For this experiment, the mixed state spread across the forward and reverse system is simulated by statistically averaging the results from two orthogonal pure single particle input states.  
We finally note that in the broken symmetry phase, Fig.~\ref{fig:fig4}(b), $\avg{S_{\ttt{F} \left(\ttt{R}\right)}}$ no longer exhibits oscillatory behaviour since the system loses its periodicity, but the final configuration of the system changes greatly depending on the initial coherence between the two subsystems.

\begin{figure*}[t!]
	\centering
	\includegraphics[trim=0 0 0 0,width=1.0\textwidth]{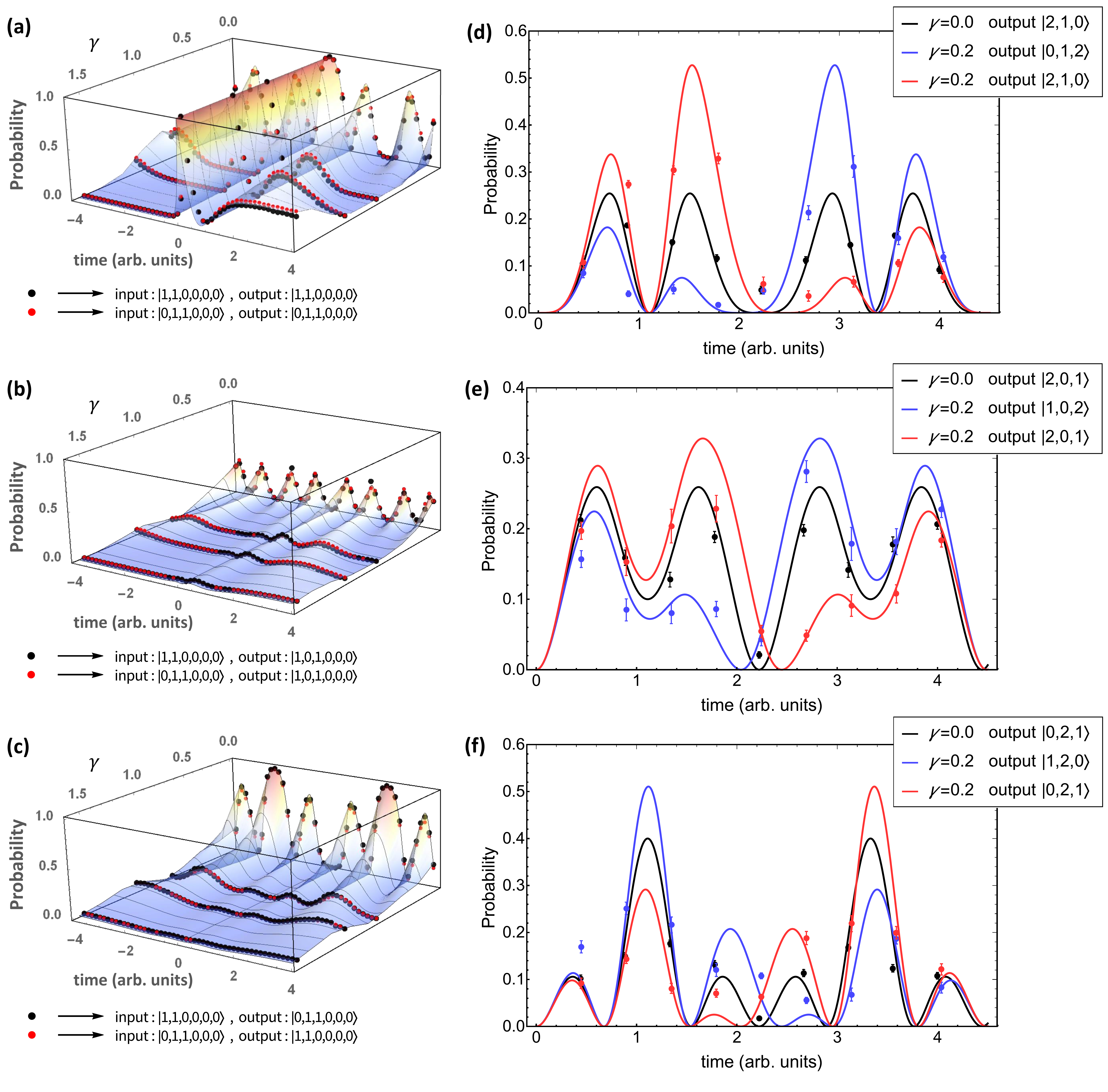}
	\caption{
	(a-c) Two-photon correlation measurements.
	Data representing the evolution of the input states
	$\ket{\Omega_1}=\ket{1_p1_p0_p}_\ttt{F}\otimes\ket{0_p0_p0_p}_\ttt{R}$
	and
	$\ket{\Omega_2}=\ket{0_p1_p1_p}_\ttt{F}\otimes\ket{0_p0_p0_p}_\ttt{R}$.
	The plot shows the probabilities of detecting the output patterns
	$\ket{1_p1_p0_p}_\ttt{F}\otimes\ket{0_p0_p0_p}_\ttt{R}$,
	$\ket{1_p0_p1_p}_\ttt{F}\otimes\ket{0_p0_p0_p}_\ttt{R}$, and
	$\ket{0_p1_p1_p}_\ttt{F}\otimes\ket{0_p0_p0_p}_\ttt{R}$,
	with dots corresponding to experimental data.
	Data collected for the state $\ket{\Omega_2}$ (in red)
	are plotted in the reverse temporal direction.
	Data for input state $\ket{\Omega_2}$ follow the same evolution
	as for input state $\ket{\Omega_1}$,
	provided that the output patterns
	$\ket{1_p1_p0_p}_\ttt{F}$ and
	$\ket{0_p1_p1_p}_\ttt{F}$ are exchanged.
	(d-f)
	Three-particle evolution. When $\gamma>0$, the output patterns related by a spatial inversion symmetry evolve in opposite directions in time. The curves for these related output patterns become identical when $\gamma=0$. Error bars are obtained via bootstrapping methods \cite{Simpson1986-bootstrap}.
	}
	\label{fig:fig5}
\end{figure*}


\subsection{Three-mode and many-particle $\pt$-symmetric systems}
We have up to now discussed the non-unitary evolution of a single excitation over 2 modes, thereby simulating single-particle $\pt$-symmetric systems.
We now fix $N=3$.
In addition to gain and loss modes, the three-mode $\pt$-symmetric systems possess a neutral mode. 
With $N=3$, the eigenvalues of $H_N(\gamma)$ become $\epsilon\in\{ 0, \sqrt{2-\gamma^2}, - \sqrt{2-\gamma^2} \} $.
We will refer to the \Appendix F for an analysis of the evolution of a single excitation over 3 modes.
Instead, here we show how the same interferometer set to realise the evolution described by  $U_{2N}(\gamma,t)$ can be used to study the interference of many identical photons undergoing non-unitary time evolution under a $\pt$-symmetric Hamiltonian. 

Using pairs of indistinguishable photons, we observe the evolution of the following two-particle input states,
\begin{equation}
\begin{split}
\ket{\Omega_1 }=\ket{1_p1_p0_p}_\ttt{F}\otimes \ket{0_p0_p0_p}_\ttt{R},\\
\ket{\Omega_2 }=\pt\ket{\Omega_1}=\ket{0_p1_p1_p}_\ttt{F}\otimes \ket{0_p0_p0_p}_\ttt{R}.
\end{split}
\end{equation}
$\ket{\Omega_1} $ has one photon in the gain and neutral modes each, while $\ket{\Omega_2} $ has one photon in the loss and neutral modes each. Figures~\ref{fig:fig5}(a-c) show the probability for the system to collapse onto gain-neutral, gain-loss, and neutral-loss modes, i.e. states $\ket{1_p1_p0_p}_\ttt{F}$, $\ket{1_p0_p1_p}_\ttt{F}$ and $\ket{0_p1_p1_p}_\ttt{F}$ respectively as a function of time. We choose four values of $\gamma$, in the Hermitian, $\pt$-symmetric ($\gamma=\sqrt{2}/2$,$\gamma=3 \sqrt{2}/4$), and   $\pt$-symmetry broken regimes ($\gamma=1.1 \sqrt{2} $). A similar procedure can be adopted to simulate the non-Hermitian two-photon (Hong-Ou-Mandel) interference~\cite{Klauck2019}.
The probabilities reported in Fig~\ref{fig:fig5}(a-c) are normalised over all the possible antibunched patterns over the six modes of the inteferometer.
We note that, due to the $\pt$ symmetry between these input states, we expect their evolution to proceed in opposite directions in time
and to present an exchange of probabilities between the observation of the patterns $\ket{1_p1_p0_p}$ and $\ket{0_p1_p1_p}$. The probability of detecting one photon in gain and neutral modes each at time $t$ starting from state $\ket{\Omega_1}$ is equal to the probability, at time $-t$, of detecting one photon in neutral and loss modes each when starting from state $\ket{\Omega_2}$. 
For these reasons, data collected for the input state $\ket{\Omega_2}$ have been plotted in the reverse temporal direction to simplify the comparison with the input state $\ket{\Omega_1}$.
By increasing the value of $\gamma$, the system undergoes the transition from the symmetric to the broken symmetry phase characterized by an increase of the period of the evolution, for $\gamma<\sqrt{2}$, and by a non-periodic evolution, for $\gamma>\sqrt{2}$. 


Lastly, we demonstrate the effect of non-hermiticity on photon-bunching by starting with a maximally anti-bunched three-photon input state,
\begin{equation} 
\ket{\chi_1}=\ket{1_p1_p1_p}_\ttt{F}\otimes \ket{0_p0_p0_p}_\ttt{R}.
\end{equation}
To prepare the desired input state, we simultaneously collect two photon pairs, then use one of these photons to herald the generation of a three-photon anti-bunched state, which we inject into the chip.
The data in Figs~\ref{fig:fig5}(d-f) show the evolution of the probability of different photon bunching events in the forward system modes. Due to the underlying parity-time symmetry, the probability distribution satisfies $\mathrm{Prob} (t,\ket{\psi})=\mathrm{Prob}(T(\gamma)-t,\mathcal{P}\ket{\psi})$ where $T(\gamma)=2\pi/\sqrt{|2-\gamma^2|}$ is the fundamental period of a $\mathcal{PT}$-symmetric trimer. Additionally, in the Hermitian limit, due to the underlying parity symmetry, the probability of two photons in mode 1 and one photon in mode 2 (i.e. pattern $\ket{2_p,1_p,0_p}_\ttt{F}$) is equal to that of its parity-symmetric pattern $\ket{0_p,1_p,2_p}_\ttt{F}$. When $\gamma>0$, the exchange symmetry between modes 1 and 3 is broken; the weight at smaller times $t<T(\gamma)/2$ shifts to the gain mode and while the weight at times $t>T(\gamma)/2$ shifts to the loss mode. For $\gamma=0.2$, parity symmetric output patterns such as $\ket{2_p0_p1_p}_\ttt{F}$ and $\ket{1_p0_p2_p}_\ttt{F}$ show probability curves that are exchanged under time-inversion $t\rightarrow T(\gamma)-t$.

It is computationally complex to simulate the evolution of quantum Hamiltonians \cite{Sparrow2018-Nature} that produce particle statistics which are governed by intractable matrix functions.
In the case of boson sampling \cite{Aaronson2013-BosonSamplingComplexity}, in which photons propagate in a circuit that is ideally described by a unitary matrix, classical intractability arises because the detection statistics are given by permanents of complex matrices.
Yet photon loss reduces the computational complexity in experimental implementations of boson sampling \cite{Renema2019-LossyBosonSampling}.
Similarly, a photonic simulation of a traditional model for an open system, such as the Lindblad approach, could become classically tractable via the loss that is necessarily built into the model.
However, in our simulation model for twin PT symmetric Hamiltonians, while each subsystem is open and lossy, the overall model is unitary and an ideal implementation is lossless;
generally, statistics provide information about both the forward and backward subsystems, and the relation between them.


\subsection{Discussion}
The theory of $\pt$ symmetric Hamiltonians emerged two decades ago as a complex extension of quantum mechanics~\cite{Bender2001}. Over the past decade, it has galvanized research primarily in the classical (wave) domain~\cite{ElGanainy2018,Miri2019}. With recent realizations of effective $\pt$-symmetry and exceptional points in minimal quantum systems, the subject of non-Hermitian quantum systems is at the forefront again. Although most studies to date have been limited to single-particle systems, non-Hermitian quantum many-particle systems are an emerging and challenging frontier. This is true because most approximate methods -- variational principle, tensor networks, perturbation theory - developed to reduce the exponential complexity of a many-body system, are designed for and work only with Hermitian Hamiltonians.

The task of simulating non-Hermitian, quantum many-particle systems, including non-interacting bosons propagating with a non-Hermitian Hamiltonian, brings new challenges \cite{Ashida2017-ManyBody,Dora2020-LuttingerLiquid}. We have shown that a programmable, unitary simulator is suited to addressing these challenges. Combined with the increasing sophistication of integrated photonic devices, our simulation techniques will allow the investigation of $\pt$-symmetric quantum systems beyond the computational capability of classical computers, paving the way for quantum technologies that can simulate and harness the potential advantages of non-Hermitian quantum systems. 

\subsection{Acknowledgments}

We acknowledge support from the Engineering and Physical Sciences Research Council (EPSRC) Hub in Quantum Computing and Simulation (EP/T001062/1). Fellowship support from EPSRC is acknowledged by A.L. (EP/N003470/1). YJ was supported by NSF grant DMR-1054020 and thanks Institute for Advanced Studies, University of Bristol for 
their support.


\subsection{Appendix A: Unitary dilation procedure}
To dilate a non-unitary operator $G_N\left(t\right)$ of size $N\times N$ we must first ensure that its maximum singular value is less than 1, $||G(t)||_2\leqslant 1$. We enforce this condition by scaling the operator $G_N\left(t\right)$ by a multiplicative factor $\alpha$. Such normalisation does not alter quantity of the form
$$
\frac{\text{Tr}\left[\Pi \alpha G(\gamma,t) \rho G^\dag(\gamma,t) \alpha^* \right]}{\text{Tr}\left[ \alpha G(\gamma,t) \rho G^\dag(\gamma,t) \alpha^* \right]} =\frac{|\alpha|^2}{|\alpha|^2} \frac{\text{Tr}\left[\Pi  G(\gamma,t) \rho G^\dag(\gamma,t)  \right]}{\text{Tr}\left[  G(\gamma,t) \rho G^\dag(\gamma,t)  \right]}
$$
where $\Pi$ is a linear operator. Therefore $\alpha$ does not affect the resulting statistics of our simulations.
We can either choose the scaling factor to be the maximum singular value over both time and spectrum of singular values, or choose to use a time varying scaling factor that is the maximum over the spectrum of singular values at time $t$. That is, $G_N\left(t\right)$ is normalised as
\begin{equation}
\Tilde{G}_N\left( t \right)  
=\frac{G_N\left(t\right)}{\max_t\left[||G_N\left( t \right) ||_2\right]}
\quad \text{or}\quad
\Tilde{G}_N\left(t\right) = \frac{G_N\left(t\right)}{\lVert G_N\left(t\right)\rVert_2}  .
\end{equation}
The second solution was preferred since it better adapts to the broken symmetry regime cases where the singular value of $G_N(t)$ increases exponentially with time. $\Tilde{G}_N\left(t\right)$ can be decomposed by singular value decomposition as $U\cos{\overrightarrow{\theta}} V^\dagger$, where $U, V$ are $N$ dimensional unitary matrices  and $\cos{\overrightarrow{\theta}} = \text{diag}\left(\cos{\theta_1}, \ldots, \cos{\theta_N}\right)$. This allows for $\Tilde{G}_N\left(t\right)$ to be embedded in the upper left corner of a $2N\times 2N$ unitary matrix. Indeed, possible dilations of $\Tilde{G}_N\left(t\right)$ can be written as
\begin{equation}
U_{2N}\left(\gamma,t\right) =
	\begin{pmatrix}
		U&&0\\0&&A
	\end{pmatrix}
	\begin{pmatrix}
		\cos{\overrightarrow{\theta}}&&i\sin{\overrightarrow{\theta}}\\i\sin{\overrightarrow{\theta}}&&\cos{\overrightarrow{\theta}}
	\end{pmatrix}
	\begin{pmatrix}
		V^\dagger&&0\\0&&B
	\end{pmatrix}
\end{equation} for any $N$ dimensional unitary matrix $A$ and $B$ and for $\sin{\overrightarrow{\theta}} = \text{diag}\left(\sqrt{1-\cos{\theta_1}^2}, \ldots,\sqrt{1- \cos{\theta_N}^2}\right)$.
Setting $A = V$ and $B=U^\dagger$ gives
\begin{equation}
U_{2N}\left(\gamma,t\right) = \begin{pmatrix}U\cos{\overrightarrow{\theta}}V^\dagger&&iU\sqrt{I-\cos^2{\overrightarrow{\theta}}}U^\dagger\\iV\sqrt{I-\cos^2{\overrightarrow{\theta}}}V^\dagger&&V\cos{\overrightarrow{\theta}}U^\dagger\end{pmatrix}
\end{equation}

Furthermore, if $C$ is a semi-positive definite diagonal matrix, by using that for a unitary matrix $W$ and a matrix $E = WCW^\dagger , \, \sqrt{E} = W\sqrt{C}W^\dagger$, we obtain 
\begin{equation}
U_{2N}\left(\gamma,t\right) = \begin{pmatrix}\Tilde{G}_N\left(t\right)&&i\sqrt{I-\Tilde{G}_N\left(t\right)\Tilde{G}_N^\dagger\left(t\right)}\\i\sqrt{I-\Tilde{G}_N^\dagger\left(t\right) \Tilde{G}_N\left(t\right)}&&\Tilde{G}_N^\dagger\left(t\right)\end{pmatrix}.
\end{equation}

Since in our model $G_N\left(t\right) = G_N^\transpose\left(t\right)$, we have $V= U^* $ and
$\left(\sqrt{I-\Tilde{G}_N^\dagger\left(t\right) \Tilde{G}_N\left(t\right) } \right) ^*= \sqrt{I-\Tilde{G}_N\left(t\right)\Tilde{G}_N^\dagger\left(t\right)}$. Then, defining the defect operator as $D_N\left(t\right)=\sqrt{I-\Tilde{G}_N\left(t\right)\Tilde{G}_N^\dagger\left(t\right)}$, we can rewrite $U_{2N}$ as
\begin{equation}
U_{2N}\left(\gamma,t\right) = \begin{pmatrix}\Tilde{G}_N\left(t\right)&&iD_N\left(t\right)\\iD_N^*\left(t\right) && \Tilde{G}_N^\dagger\left(t\right)\end{pmatrix}
\end{equation}

In the case when $H_N$ is Hermitian, $G_N\left(t\right)$ is unitary and the singular values are $1$, so $D_N = 0$, giving two independently evolving systems.
\begin{figure*}
	\centering
	\includegraphics[width=\textwidth]{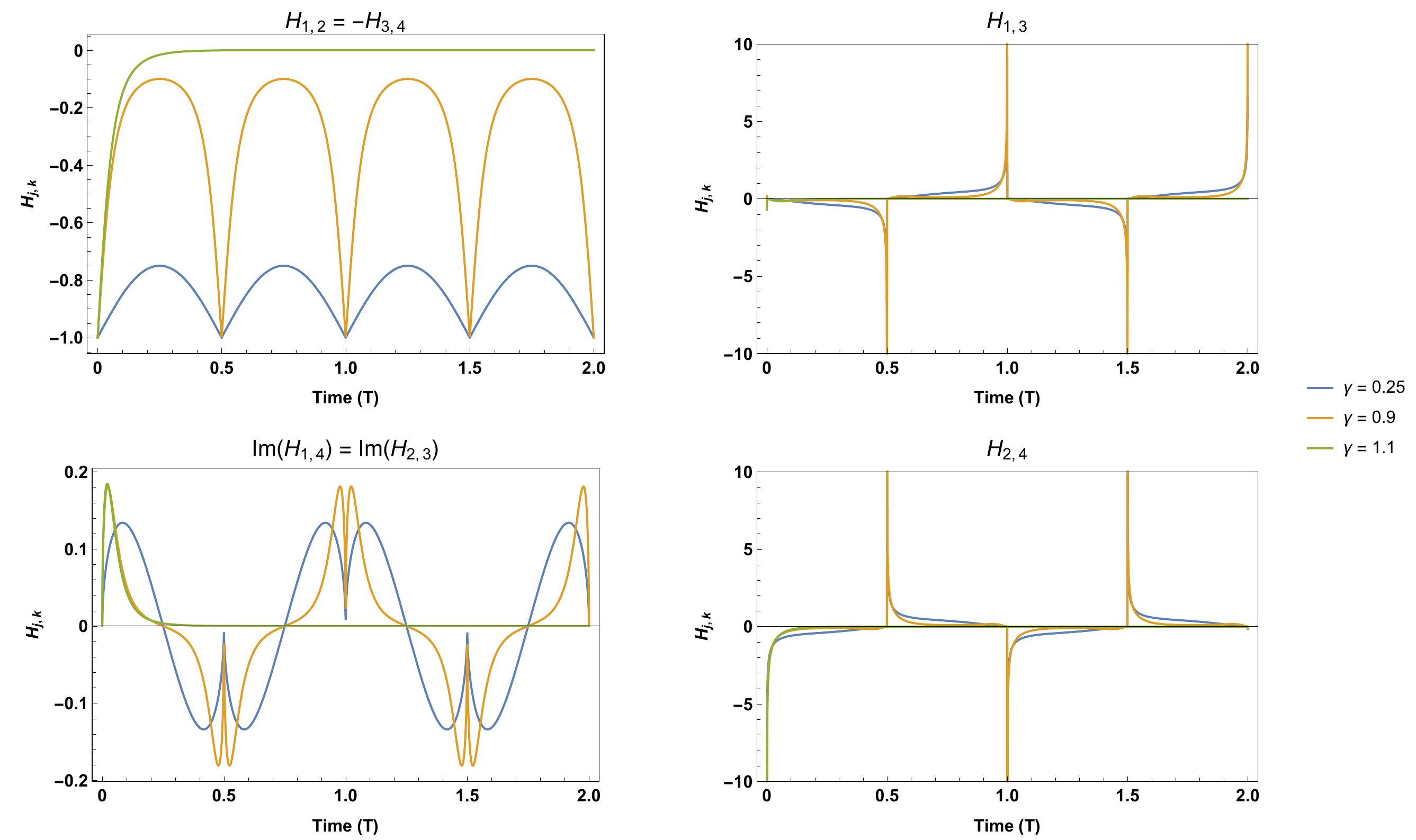}
	\caption{non-zero elements of the effective Hamiltonian $H_{\text{eff}}$ of the dilated unitary for the $2-$ mode $\pt$-symmetric Hamiltonian $H_{2}$}
	\label{fig:EffectiveHamiltonian}
\end{figure*}
\subsection{Appendix B: Effective Hamiltonian}
As the evolution of the pair of coupled systems is unitary, it can be generated by a Hermitian Hamiltonian $H_\text{eff}$ over the $2N$ modes satisfying the equation 
\begin{equation}
H_\text{eff}U_{2N}=i \frac{\text{d}}{\text{d} t}U_{2N}.
\end{equation}
From this, knowing $U_{2N}$ it is possible to obtain $H_{\text{eff}} = i\left(\frac{d}{dt}U_{2N}\right)U_{2N}^{\dagger}$ and numerically solve for $H_{\text{eff}}$, which turns out to be time dependent. In Fig. \ref{fig:EffectiveHamiltonian} we show the results of the numerical calculation of the $H_\text{eff}$ elements for $N=2$ and $\gamma \in \{0.25, 0.9, 1.1\}$. In the symmetric regime, the elements oscillate with a recurrence time $\tau=2\pi/\sqrt{1-\gamma^2}$ so we set an equivalent `relaxation time' $\tau=2\pi/\sqrt{|1-\gamma^2|}$ in the broken regime. The unplotted elements in figure are all (numerically) $0$ including all diagonal elements $H_{k,k}$ for $1\leq k\geq 4$. The spikes appearing in $H_{1, 3}$ and $H_{2, 4}$ represent asymptotes.


\subsection{Appendix C: Forward and reverse systems with vacuum contribution}
The dynamics of the forward (F) and reverse (R) systems are recovered by the reduced states of the full $2N$-mode system:
\begin{equation}
\begin{split}
\rho_\ttt{F}(t) &= \ttt{Tr}_\ttt{R}\, \rho(t):=\ttt{Tr}_\ttt{R} \big[ \mathcal{U}_{2N}(\gamma,t) \rho(0) \mathcal{U}_{2N}^{\dag}(\gamma,t) \big],  \\
\rho_\ttt{R}(t) &= \ttt{Tr}_\ttt{F}\, \rho(t):=\ttt{Tr}_\ttt{F} \big[ \mathcal{U}_{2N}(\gamma,t) \rho(0) \mathcal{U}_{2N}^{\dag}(\gamma,t) \big].
\end{split}
\end{equation}
Here, $\rho(0)$ is the input photonic state and $\mathcal{U}_{2N}$ is the unitary operator acting on the photonic Fock space according to the transfer matrix $U_{2N}(\gamma,t) $, which transforms the input-mode creation operators $a_i^\dagger$ as
\begin{equation}
 \mathcal{U}_{2N}(\gamma,t)  a_i^{\dag} \mathcal{U}^{\dag}_{2N}(\gamma,t)  = \sum_j [U_{2N}(\gamma,t) ]_{ji} a^{\dag}_j.
\end{equation}

To simulate a single $N$-dimensional $\pt$ system, we use $2N$ optical modes and we assume that the optical system is constrained to the one-photon subspace. 
In this space a basis is

\begin{equation} \label{eq:genericState}
 \left\lbrace\left|k\right> = \left|0_p\right>^{\otimes \left(k-1\right)}\left|1_p\right>\left|0_p\right>^{\otimes \left(2N-k\right)}\right\rbrace_{k=1}^{2N}
\end{equation}
where $\left|0_p\right> $ is the single mode vacuum state. Doing this, the density matrix at time $t$ can then be denoted as
\begin{equation}
 \rho\left(t\right) =U_{2N}\left(\gamma,t\right) \rho\left(0\right)U_{2N}^\dag \left(\gamma,t\right) .
\end{equation}
However we can treat the modes $1$ to $N$ as composing the forward system and modes from $N+1$ to $2N$ as composing the reverse system leading to the definition of two reduced density matrices obtained performing the partial trace over one of the systems. For the reduced density matrix of the forward system $\rho_F$ we obtain:
\begin{widetext}
\begin{align*}
\rho_F = \mathrm{tr}_R\rho &= \mathrm{tr}_R \sum_{k, l = 1}^{2N}p_{kl}\left|k\rangle\langle l\right| \\
&=\mathrm{tr}_R \sum_{k, l = 1}^{2N}p_{kl} \left|0_p\right>^{\otimes \left(k-1\right)}\left|1_p\right>\left|0_p\right>^{\otimes\left(2N-k\right) }\left<0_p\right|^{\otimes \left(l-1\right)}\left<1_p\right|\left<0_p\right|^{\otimes\left(2N-l\right)} \\
&= \mathrm{tr}_R \biggl(\sum_{k, l = 1}^N p_{kl} \left|0_p\right>^{\otimes \left(k-1\right)}\left|1_p\right>\left|0_p\right>^{\otimes\left(2N-k\right) }\left<0_p\right|^{\otimes \left(l-1\right)}\left<1_p\right|\left<0_p\right|^{\otimes\left(2N-l\right)} \\
&\qquad\qquad + \sum_{k, l = N+1}^{2N} p_{kl} \left|0_p\right>^{\otimes \left(k-1\right)}\left|1_p\right>\left|0_p\right>^{\otimes\left(2N-k\right) }\left<0_p\right|^{\otimes \left(l-1\right)}\left<1_p\right|\left<0_p\right|^{\otimes\left(2N-l\right)} \\
&\qquad\qquad + \sum_{k=1}^N\sum_{l = N+1}^{2N} p_{kl} \left|0_p\right>^{\otimes \left(k-1\right)}\left|1_p\right>\left|0_p\right>^{\otimes\left(2N-k\right) }\left<0_p\right|^{\otimes \left(l-1\right)}\left<1_p\right|\left<0_p\right|^{\otimes\left(2N-l\right)}\\
&\qquad\qquad + \sum_{k= N+1}^{2N}\sum_{l=1}^N p_{kl} \left|0_p\right>^{\otimes \left(k-1\right)}\left|1_p\right>\left|0_p\right>^{\otimes\left(2N-k\right) }\left<0_p\right|^{\otimes \left(l-1\right)}\left<1_p\right|\left<0_p\right|^{\otimes\left(2N-l\right)}\biggr) \\
&= \sum_{k, l = 1}^N p_{kl} \left|0_p\right>^{\otimes \left(k-1\right)}\left|1_p\right>\left|0_p\right>^{\otimes\left(N-k\right) }\left<0_p\right|^{\otimes \left(l-1\right)}\left<1_p\right|\left<0_p\right|^{\otimes\left(N-l\right)} + \sum_{k, l = N+1}^{2N}p_{kl}\left|0_p\right>^{\otimes N}\left<0_p\right|^{\otimes N}\delta_{k, l} \\
&= \sum_{k, l = 1}^N p_{kl} \left|k\rangle\langle l\right| + \sum_{k = N+1}^{2N}p_{kk}\left|\text{vac}\rangle\langle\text{vac}\right|
\end{align*}
\end{widetext}
Similar results are obtained when the partial trace is performed over the forward system, showing that the partial trace operation over the reverse (forward) system sums the diagonal elements of the bottom (top) $N$ modes of $\rho\left(t\right)$ and places this value in the vacuum state of the forward (reverse) system. That is, the Fock spaces of the systems are decomposed into particle number subspaces as $\mathcal{F}^{\left(F/R\right)} = \mathcal{H}^{\left(F/R\right)}_0\oplus\mathcal{H}^{\left(F/R\right)}_1$. According to this, we can choose the basis states of the forward or reverse system as $\left\lbrace\left|k\right>^{\left(F/R\right)} = \left|0_p\right>^{\otimes \left(k-1\right)}\left|1_p\right>\left|0_p\right>^{\otimes \left(N-k\right)}\right\rbrace_{k=1}^N$ and an additional basis state in each system $\ketNull=\left|0_p\right>^{\otimes N} $.
Furthermore, no coherence is present between the 1 photon and 0 photon subspaces.

Whenever there is interest in considering only the original $\pt$ system without introducing the reverse system and the interaction terms between them, we need to redefine the density matrix of the forward system without including the vacuum contribution as
\begin{equation}
 \Tilde{\rho}_F = \frac{\sum_{k, l = 1}^N p_{kl} \left|k\rangle\langle l\right|}{\sum_{k=1}^N p_{kk}}
\end{equation}
This approach has been used in the simulation of the no-signalling violation.

\subsection{Appendix D: Experimental methods}
\vspace{-0.2 cm}
\subsubsection{\textbf{\emph{Photon generation}}}
\vspace{-0.5 cm}
For all the experiments described in this paper, the single photons are generated via spontaneous parametric down-conversion 
in a Bismuth Borate non-linear crystal satisfying type I phase-matching conditions. The crystal is pumped with light pulses at 404~nm obtained from a second harmonic generation process induced in a Barium Borate crystal by a 808~nm Ti:Sapphire mode-locked laser. The pairs of photons generated in the down-conversion are spectrally filtered with a 3~nm bandwidth interferometric filter centred around 808~nm and are collected from two diametrically opposite positions of the generation cone. The pair of photons, coupled to single mode polarisation maintaining fibres, are used either as a two-photon Fock state, by injecting them both into our interferometer, or as heralded single photon by connecting one of the two fibres to a single photon detector and injecting the other photon into the interferometer. When interested in the two-photon Fock state we record all the events where two single photons are detected at the output of the chip. For the experiments in which we only need a single photon source, we record all the events when both the heralding photon and the photon at the output of the chip are detected. Three-photon experiments are performed collecting two separate pairs of photons from the emission cone and recording the coincidental detection of a heralding photon from one pair and of three photons passing through the optical modes of the interferometer.

\vspace{-0.2 cm}
\subsubsection{\textbf{\emph{Integrated interferometer}}}
\vspace{-0.5 cm}
The programmable interferometer used to implement the multiple unitary transfer matrices of the experiments was produced by the NTTGroup in Japan and is described in \cite{carolan2015-universal}. The triangular arrangement of 15 integrated Mach-Zehnder interferometers with 30 thermally tunable phase-shifters allows us to prepare any unitary optical transfer matrix across 6 modes.  Alternatively, the same device can be used as a cascade of $2\times2$ interferometers in order to sequentially implement state preparation, evolution or projection in physically different parts of the chip.

\vspace{-0.2 cm}
\subsubsection{\textbf{\emph{Probabilistic number resolving}}}
\vspace{-0.5 cm}
Three photons simulations reported in this article require the ability to resolve multiple occupancy in the output modes to detect bunching photon events. To measure the probability of these events, we performed probabilistic number resolving detection of up to two photons in the same mode using auxiliary fibre beamsplitters (FBS) and detectors. By inserting a FBS at each of the first 3 output modes of the inteferometer and connecting both output modes of each FBS to detectors, there is a finite probability that photons bunched in the same output mode separate at the FBS and generate a signal at both detectors to which the FBS is connected.
\begin{figure*}
	\centering
	\includegraphics[trim=0 0 0 0,width=1.0\textwidth]{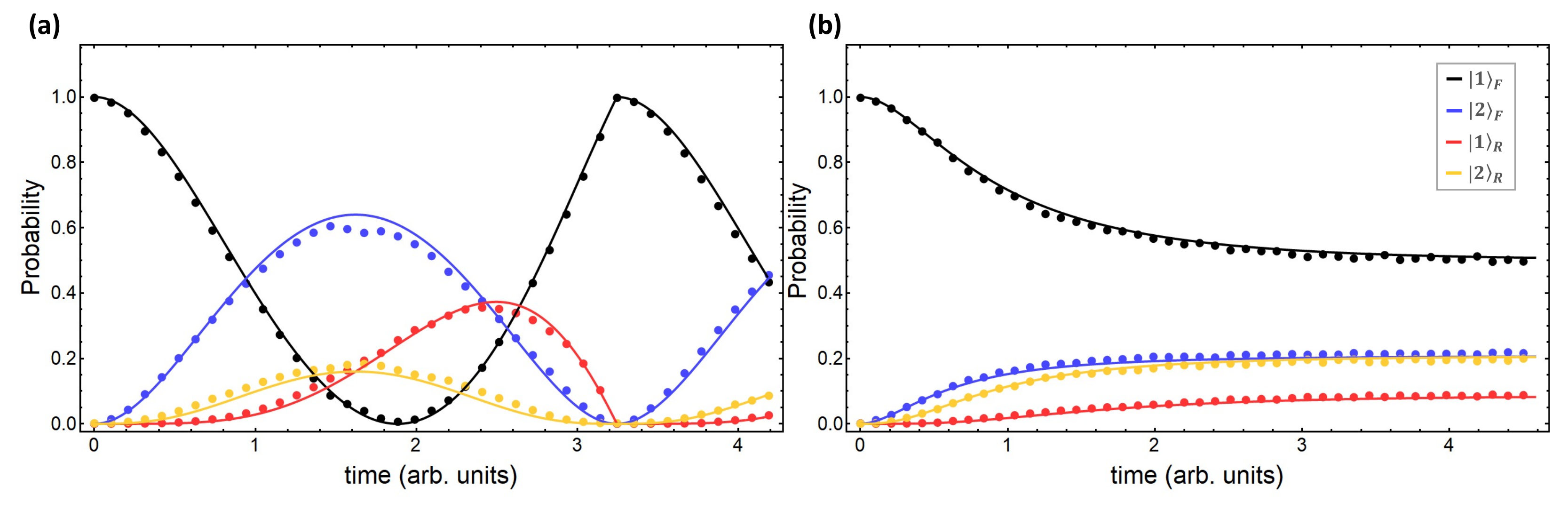}
	\caption{Plots showing the evolution of a single excitation in the forward subsystem. Solid lines (points) show the theoretical curve (experimental data). We show the probabilities for the excitation being detected in the forward 
	subsystem ($\ket{1}_\mathrm{F}$, $\ket{2}_\mathrm{F}$) and the reverse subsystem ($\ket{1}_\mathrm{R}$, $\ket{2}_\mathrm{R}$). Both panels follow the legend in (b). (a) Plots for a $\pt$-Hamiltonian in the unbroken regime ($\gamma = 0.25$). We see periodic  oscillations of the excitation in all four modes. 
	(b) Plots for the broken symmetry regime. All probabilities evolve exponentially to a steady state ($\gamma = 1.1$).}
	\label{fig:fig_app_E}
\end{figure*}
%
\vspace{-0.2 cm}
\subsubsection{\textbf{\emph{Detection calibration}}}
\vspace{-0.5 cm}
We experimentally estimate the probability of collapsing the quantum state of our simulator onto a particulat Fock state by multiplying the number of occurrence of each detected pattern by a detector calibration factor associated to that pattern, and then normalising to unity the sum over all relevent output patterns. 
The relative detector efficiency at the different output modes, for single photon events, is obtained by testing 500 random phases configurations of our interferometer and comparing the experimental statistic with the expected photon statistics due to the phase settings. The efficiency corrections of each mode that maximises the statistical fidelity between the experimental statistics and the theoretical distribution is chosen as numerical correction for the data collected during single photon experiments. 
For two-photon anti-bunched events, the occurence of each pattern is multiplied by the product of the correction factor of the individual output modes populated with photons. 
 
In the pseudo-number-resolving configuration, the detection calibration factor is obtained by using the single photon detector efficiency $\nu_i$ recorded at each output of the FBS. To take into account the probabilistic nature of the pseudo-number resolving scheme, the experimental occurrence of a detection pattern such as 
$
((n_{1,1}, n_{1,2}), (n_{2,1}, n_{2,2}), (n_{3,1}, n_{3,2}))
$ is divided by   
\begin{multline}
   \Gamma((n_{1,1}, n_{1,2}), (n_{2,1}, n_{2,2}), (n_{3,1}, n_{3,2})) \\= \prod_{i=1}^{i=3}  \nu_{i,1}^{n_{i,1}}\nu_{i,2}^{n_{i,2}}  (n_{i,1}+n_{i,2})!
\end{multline}
The number of events associated to an output state $\ket{N_{1p} N_{2p} N_{3p}}$ with $N_{kp}=n_{k,1}+n_{k,2}$ is given as the average of the occurrences of equivalent detection patterns. For example, 
	$((1, 1), (1, 0), (0, 0))$  is equivalent to $((1, 1), ( 0,1), (0, 0))$ since both correspond to the output state $\ket{2_p 1_p 0_p}$.
The final probabilities are normalised over all the output states characterised by three photons in the forward system modes but with less than three photons bunched in the same mode.

\vspace{-0.2 cm}
\subsubsection{\textbf{\emph{State preparation and reconstruction}}}
\vspace{-0.5 cm}
The preparation of fully mixed single photon states is obtained by interfering two indistinguishable single photons. The Hong Ou Mandel visibility of their inteference is measured to be 0.97. During the entropy evolution simulation, the inteferometer is set to implement $U_{\ttt{Proj}}\ U_{4}\left(\gamma,t\right) \ U_{\ttt{Ent}}$, where :  
\begin{equation}
	U_{\ttt{Ent}} = \begin{pmatrix} 0 & 0 & \frac{\sqrt{2}}{2} & \frac{\sqrt{2}}{2} & 0 & 0 \\ 
					0 & 0  & 0 & 0& \frac{\sqrt{2}}{2} & \frac{\sqrt{2}}{2} \\ 
					1 & 0 & 0 & 0 & 0 & 0 \\ 
					0 & 1 & 0 & 0 & 0 & 0 \\ 
					0 & 0   & \frac{\sqrt{2}}{2} & -\frac{\sqrt{2}}{2} & 0 & 0\\ 
					0 & 0 & 0 & 0 & \frac{\sqrt{2}}{2} & -\frac{\sqrt{2}}{2}  \\ 
    \end{pmatrix},
\end{equation}
 $U_{4}\left(\gamma,t\right)$ transforms the bottom four modes of the interferometer, and $U_{\ttt{Proj}}$ projects the bottom two modes on the bases of the Pauli operators. Physically separate components of the optical circuit are involved in the implementation of the three cascaded unitaries. 
 \begin{figure*}[t]
	\centering
	\includegraphics[trim=0 0 0 0,width=1.0\textwidth]{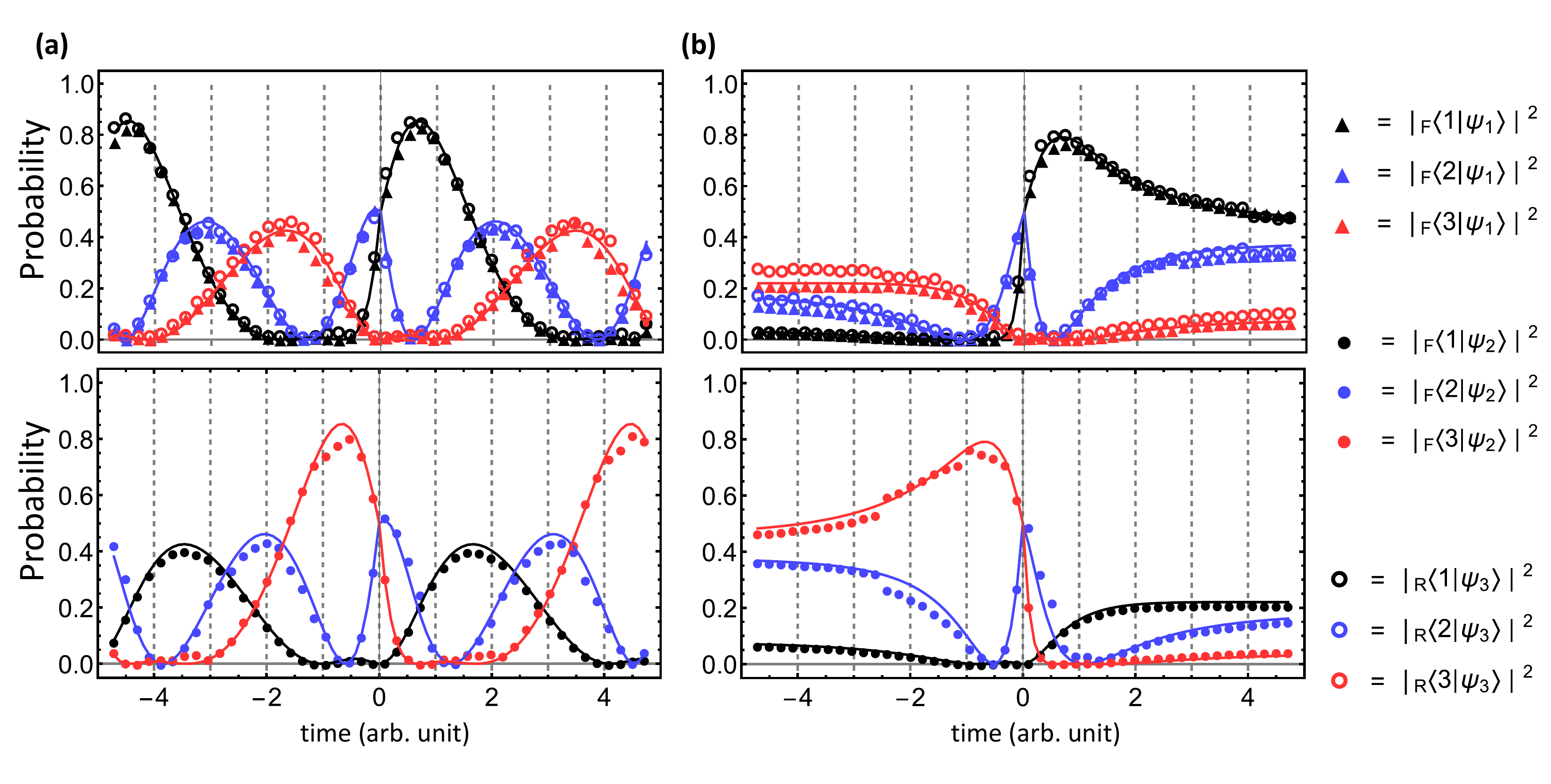}
	\caption{
	Results for the simulation of the 3-level $\pt$-symmetric Hamiltonian.
	Plots show evolution of the probability, in the unbroken (a) ($\gamma=0.5 \gammaC$) and broken (b) ($\gamma=1.1 \gammaC$) symmetry regimes,
	for three input states:
	$ \ket{\psi_1(t=0)}=(\ket{1}_\ttt{F}-i\ket{2}_\ttt{F})/\sqrt{2}$,
	$ \ket{\psi_3(t=0)}=(\ket{1}_\ttt{R}+i \ket{2}_\ttt{R})/\sqrt{2}$ (top row), and
	$\ket{\psi_2(t=0)}=(i\ket{2}_\ttt{F}+\ket{3}_\ttt{F})/\sqrt{2} $ (bottom row).
	Solid lines in the top rows represent theory curves for the time evolution of $\ket{\psi_1(t)}$
	while solid lines in the bottom rows represent theory curves for the time evolution of $\ket{\psi_2(t)}$.
	The data for
	$\ket{\psi_3}$ show the same temporal behaviour as for $\ket{\psi_1}$.
	Conversely, the data for $\ket{\psi_2}$ follow the opposite temporal evolution as
	$\ket{\psi_1}$, while the probabilities of observing $\ket{1}$ and $\ket{3}$ are exchanged.
	}
	\label{fig:fig3}
\end{figure*}
The evolution of excitations superposed across the forward and reverse systems is performed by compiling the two transformations required for preparing and evolving the states. The resulting unitary is implemented in the first part of the chip. The expectation values of $S_F$ and $S_R$ are measured using 3 further concatenated interferometers to project each of the systems, independently, on the eigenstates of the Pauli operators.

When $N=3$, to simulate the evolution of a single particle state prepared in a superposition across multiple modes of the forward or reverse system, we multiplied the transfer matrix   $U_{6}\left(\gamma,t\right)$ with the state rotation necessary to transform a localised single photon into the desired superposition. The resulting unitary transformation is then dialed on the optical chip.

The experiments with two-photon Fock states are performed setting the transfer matrix of the chip to implement $U_{6}(\gamma,t)$ while the input state is obtained injecting either $\ket{0_p1_p1_p}_\ttt{F}\otimes \ket{0_p0_p0_p}_\ttt{R}$ or $\ket{1_p1_p0_p}_\ttt{F}\otimes\ket{0_p0_p0_p}_\ttt{R}$ into our integrated interferometer. The production of the desired input state is assumed everytime a coincidence detection happens at two of the detectors connected to the output modes of our device neglecting higher order pair generation occurring with a two order of magnitude lower rate.



\subsection{Appendix E: Two-mode system: reverse subsystem}

In the main text we show the dynamics of an excitation in the forward subsystem for a two-mode $\pt$-system. We included a combined vacuum probability which contained the 
probability of the excitation tunnelling into the reverse subsystem. Using the same experimental data, in Fig.~\ref{fig:fig_app_E} we show the evolution of both forward and 
both reverse modes. This captures the full dynamics of a single excitation in the forward subsystem. We show the data for both the unbroken ($\gamma = 0.25$) and broken 
($\gamma = 1.1$) symmetry regimes. We see a periodic oscillation in the probability of all four modes in the unbroken regime. In the broken regime all four mode probabilities 
evolve exponentially towards a steady state value.

\subsection{Appendix F: Three-mode system: single particle dynamics}

Since, for $N=3$, the third-order exceptional point occurs at $\gammaC=\sqrt{2}$, to analyse unbroken and broken symmetry regimes
we perform simulations 
for $\gamma=\sqrt{2}/2$
and $\gamma=1.1\,\sqrt{2}$, respectively.
Transfer matrices, concatenating state preparation and unitary evolution described by $U_6(\gamma,t_{i})$, are implemented by using the entirety of the six mode chip.

Figure~\ref{fig:fig3} shows data reproducing the evolution of three single-particle initial states
\{$\ket{\psi_1 (t=0)}_\ttt{F}$, $\ket{\psi_2(t=0)}_\ttt{F}$, $\ket{\psi_3(t=0)}_\ttt{R}$\}.
These are related by symmetry transformations as follows:
\begin{equation}
\ket{\psi_1(0)}_\ttt{F}=(\ket{1}_\ttt{F}-i\ket{2}_\ttt{F})/\sqrt{2}
\end{equation}
is an example of a state coherently superposed across gain and neutral sites of the forward system;
\begin{align}
\ket{\psi_2(0)}_\ttt{F}& =(i\ket{2}_\ttt{F}+\ket{3}_\ttt{F})/\sqrt{2}=\pt \ket{\psi_1(0)}_\ttt{F}
\end{align}
is the result of a $\pt$ transformation of $\ket{\psi_1(0)}_\ttt{F}$;
\begin{align}
\ket{\psi_3(0)}_\ttt{R}& =(\ket{1}_\ttt{R}+i \ket{2}_\ttt{R})/\sqrt{2}=\mathcal{T} \ket{\psi_1(0)}_\ttt{R}
\end{align}
is prepared in the reverse system and, via time inversion, is analogous to the state $\mathcal{T} \ket{\psi_1(0)}_\ttt{F}$.

Due to the symmetries of our model, the mode occupation of these states during their evolution is closely related. Specifically, we expect:
\begin{align}
\braket{k}{\psi_2(t)} &= \braket{k}{\pt G_3(-t) \pt \psi_2(0)} \nonumber \\
&= \braket{4-k}{G_3(-t)\psi_1(0)}^*,
\end{align}
and, since $H_\text{3}(\gamma)=H_\text{3}^\ttt{T}(\gamma)$, we also have
\begin{align}
\braket{k}{G_3(t)^\dag \psi_3(0)} &= \braket{k}{\mathcal{T}  G_3(t) \mathcal{T}  \psi_3(0)} \nonumber \\ 
&= \braket{k}{\mathcal{T} G_3(t)\psi_1(0)}  \nonumber \\
&= \braket{k}{ G_3(t)\psi_1(0)}^*
\end{align}

Figures~\ref{fig:fig3}(a,b) show that the (forward) evolution of state $\ket{\psi_1}$ with Hamiltonian $H_3$ is the same as the (reverse) evolution of the state $\ket{\psi_3}$ with Hamiltonian $H_3^*$. These results at time $-t$ match the evolution of state $\ket{\psi_2}$ with Hamiltonian $H_3$ at time $t$, demonstrating the mirror symmetry about $t=0$. 

\bibliography{pt_polished}

\begin{thebibliography}{10}
\bibitem{foot1}
For continuous-variable systems, the coordinate operator is defined as
  $\hat{x}=(\hat{a}^\dagger+\hat{a})/\sqrt{2}$ where $\hat{a}^\dagger$ and
  $\hat{a}$ are the particle creation and annihilation operators. In turn, for
  discrete-variable systems, one can write down Pauli $\hat{\sigma}^x$ matrix
  as $\hat{\sigma}^x=(\hat{S}^++\hat{S}^-)/\sqrt{2}$, where $\hat{S}^{\pm}$ are
  Pauli ladder operators. Since the creation and annihilation operators are
  similar to ladder operators in their action in the state space, in this
  context, we consider $\hat{\sigma}^x$ as the discrete-variable analog of
  $\hat{x}$.
\end{thebibliography}
\bibliographystyle{ieeetr}

\end{document}